# SUePDF: a program to obtain quantitative pair distribution function from electron diffraction data

Dung Trung Tran, Gunnar Svensson and Cheuk-Wai Tai

Materials and Environmental Chemistry, Arrhenius Laboratory,

Stockholm University, Stockholm, S-10691, Sweden

**Abstract**

SUePDF is a graphic-user-interface program written in MATLAB to achieve quantitative pair distribution functions (PDF) from electron diffraction data. The program facilitates the structural studies of amorphous materials and small nanoparticles based on electron diffraction data from transmission electron microscopes (TEMs). It is based on the physics of electron scattering as well as the total scattering methodology. A method of background modelling is introduced to treat the intensity tail of the direct beam, inelastic scattering and incoherent multiple scattering. Kinematical electron scattering intensity is scaled using the electron scattering factors. The PDFs obtained after Fourier transforms are normalized with respect to number density, nanoparticle form factor, and the non-negativity of probability density. SUePDF is distributed as free software for academic users.

Correspondence email: dung.tran@mmk.su.se; cheuk-wai.tai@mmk.su.se

# 1. Introduction

The PDF method is widely employed for studying structurally disordered materials (Warren, 1990; Egami & Billinge, 2002; Proffen et al., 2003). This is because it gives more structural information, beyond the information given by the traditional Bragg-peak-based analysis, from scattering data. Standard computer programs for PDFs for neutron and X-ray powder diffraction data are well established (Peterson et al., 2000; Juhás et al., 2013). There have been made a number of efforts to obtain PDFs from electron diffraction (ED) data (Cockayne & McKenzie, 1988; Tewes et al., 1994; Takagi et al., 2001; McBride & Cockayne, 2003; Ankele et al., 2005; Ishimaru et al., 2008; Abeykoon et al., 2012; Mu et al., 2013), however, these have not been sufficient to establish electron powder diffraction as one of the major data sources for PDF analysis. The main obstacle has been the multiple scattering of electrons, which may alter the scattering intensities in the ED patterns making it difficult to extract an undistorted structure function (Uyeda, 1968; Cowley, 1969; Anstis et al., 1988; Cockayne & McKenzie, 1988). Beside this, the electron-matter interaction differs from the cases of with X-rays and neutrons, implying that a dedicated procedure for of data treatment is needed for ED data.

In order to obtain reasonable PDFs, ED data $I(Q)$ has to be scaled using appropriate electron scattering factors $f_e(Q)$ to yield a proper reduced structure function $F(Q)$, where $Q$ is the magnitude of the scattering vector. This is not an easy task because the ED data is generally stained by inelastic scattering and distorted by multiple scattering, instrument errors and noises. The widely mentioned solution for inelastic scattering is using an energy filter but this may also cause additional distortion in the data when only electrons of certain energy-loss range can be blocked. Data scaling was previously carried out by multiplying the compositional averages $\langle f_e^2(Q) \rangle$ and $\langle f_e(Q) \rangle^2$ with a fitting parameter $\eta$ which may be fixed by matching $\eta \langle f_e^2(Q) \rangle$ with $I(Q)$ at large $Q$ (Cockayne & McKenzie, 1988; Cockayne, 2007; Cockayne et al., 2010). An early software attempt to fit $I(Q)$ with $\langle f_e^2(Q) \rangle$ for PDF extraction was made by drawing and then subtracting a background of $I(Q)$ (Hauschild, 2009). However, because the mathematical curve presented for this background is generally not suitable, the background is often made manually. The next software attempt was made by Mitchell and Petersen (2012), in which an additional fitting parameter $\alpha$ is introduced to counter the discrepancy between $I(Q)$ and $\eta \langle f_e^2(Q) \rangle$. This means, $F(Q)$ is now obtained by matching $\eta \langle f_e^2(Q) \rangle - \alpha$ with $I(Q)$ at large $Q$. In other work, instead of introducing an additional parameter, $F(Q)$ was directly adjusted by fitting the distortion feature to a fourth-order polynomial (Mu *et al.*, 2013).

The program SUePDF aims to effectively correct the ED data by taking into account the multiple scattering features and uniqueness of electron-matter interactions. For amorphous materials, the coherent multiple scattering can be considered as insignificant and incoherent multiple scattering should merely contribute to a background (Cowley, 1992), which may be modeled and subtracted. For crystalline-like nanoparticles, coherent multiple scattering should not affect the peak positions of the resulting PDFs but only the peak intensities. This has been demonstrated to be the case for crystals having their thickness less than five times of the electron mean free path (Anstis et al., 1988). The problem with modulated peak intensities caused by coherent multiple scattering is tackled in SUePDF by a renormalization procedure based on number densities and the probability of having an atom at a certain distance is non-negative. In general, the major distortion of $I(Q)$ can be considered to form a smooth background, which is built up from the direct-beam tail, inelastic continuum, and incoherent multiple scattering. SUePDF employs the combination of an optimizing parameterization and a reasonable mathematical model for the background. Inside SUePDF, background subtraction is coupled with data scaling in a loop-based routine to optimize the data treatment. After background subtraction and data scaling, the PDF can be obtained by a Fourier transform of the normalized data. On the basis of the physical meaning of PDFs, noise filtering and normalization procedures have to be routinely carried out. This improves the physical reliability of the outcome PDFs and allows the uncertainties to be evaluated. SUePDF also offers the possibility to correct finite size effects present for nanoparticles, by using a *nanoparticle form factor* computed for a given size and shape.

## 2. Methods

### 2.1. Background modelling

The smooth background observed for ED data is considered to be the contributions from the direct-beam tail, inelastic scattering continuum and incoherent multiple scattering. For an elastic scattering of a $\lambda$-wavelength electron with an semi-angle $\theta$, $Q = 4\pi \sin\theta / \lambda$, but in the case of small-angle inelastic scattering (Egerton, 2011):

$$Q = \sqrt{(\frac{4\pi \sin\theta}{\lambda})^2 + (\frac{2\pi \bar{\theta}}{\lambda})^2} \qquad (1)$$

where $\bar{\theta} = \gamma_E \Delta E / [K_E(1+\gamma_E)]$ is the characteristic angle corresponding to an energy loss $\Delta E$; $\gamma_E = (1 + K_E / E_0)$ is the relativistic factor with $K_E$ as the kinetic energy and $E_0 \approx 511\ keV$ as the stationary mass-converted energy of the electron, respectively. Eq. (2) can be rewritten for the inelastic component when $\theta = 0$:

$$Q_{inel}(\Delta E) = \frac{2\pi \gamma_E \Delta E}{\lambda K_E (1+\gamma_E)} \qquad (2)$$

For example, if the electron energy loss is counted up to 2000 eV for an incident beam of 200 keV, then the inelastic beam tail is limited to below $Q_{inel}(2000) \approx 1.46$ Å$^{-1}$. For low-loss electrons ($\Delta E < 50$ eV), the inelastic component $Q_{inel}(<50)$ is less than ~0.036 Å$^{-1}$. Eq. (2) means that the inelastic error is more significant for electrons of high energy-loss. It is well-known that, high-loss electrons can be excluded using an energy filter. However, according to our experience, energy filtering also modifies the background, making an accurate quantitative background modeling difficult. The following mathematical model is introduced to fit the background $B(Q)$ of the electron powder diffraction pattern:

$$B_N(Q) = \sum_{k=1}^{N} \frac{c_k}{Q^k} \qquad (3)$$

Eq. (3) is actually a positive-degree portion of a Laurent-type series where $c_k$ are the fitting parameters and $N$ is the fitting order which may vary for different sample composition. A comparison between the power-law model and the Laurent-type model [described in Eq. (3), with $N = 7$] for background fitting of an electron powder diffraction data acquired from a nanoporous carbon sample is shown in Fig. (1). In addition, this background model works in both thin and thick sample. It is demonstrated in a study of amorphous silica discussed in Section 4.2. The measured bond-length and the values reported in literature are listed and are in good agreement.

## 2.2. Optimization procedure for data scaling and background modeling

Electron scattering factors, denoted as $f_e(Q)$, are used for data scaling. These factors may be obtained using the following Mott-Bethe formula (Mott & Massey, 1965):

$$f_e(Q) = \frac{2\gamma}{a_0}\left\{\frac{Z - V - f_X(Q)}{Q^2}\right\} \qquad (4)$$

where $V$ is the valence number, $a_0 \approx 0.53$ Å is the Bohr radius and $f_X(Q)$ is the X-ray scattering factors (Brown et al., 2006). For neutral atoms, it is recommended to use the available DFT-computation-based parameterizations (Kirkland, 2010), which are considered to be more accurate than those from the Mott-Bethe formula, particularly at low-Q values. For samples composed of more than one element, the chemical compositions are required as input together with the corresponding valence for each element.

The data scaling is constrained by a mathematical feature of the structure function $S(Q)$. That is $S(Q) \rightarrow 1$ when $Q \rightarrow \infty$. The background-subtracted and normalized scattering intensity $I(Q)$ is converted into $S(Q)$ by using the composition-averaged $\langle f_e^2(Q) \rangle$ and $\langle f_e(Q) \rangle^2$ (Warren, 1990):

$$S(Q) = 1 + \frac{I(Q) - \langle f_e^2(Q) \rangle}{\langle f_e(Q) \rangle^2} \qquad (5)$$

The data scaling is done according to:

$$I(Q) = \frac{\int_{Q_{min}}^{Q_{max}} \langle f_e^2(Q') \rangle dQ'}{\int_{Q_{min}}^{Q_{max}} [I_{raw}(Q') - B_N(Q')] dQ'} [I_{raw}(Q) - B_N(Q)] \qquad (6)$$

where $I_{raw}(Q)$ is the raw data of scattering intensity. The feature $S(Q) \to 1$ when $Q \to \infty$ restricts $I(Q)$ to attenuate around $\langle f_e^2(Q) \rangle$ at high Q values. This suggests a necessary minimization of the following quantity, defined at the tail of $I(Q)$:

$$\chi_{tail}^2 = \int_{tail} \left[ I_{tail}(Q) - \langle f_e^2(Q) \rangle_{tail} \right]^2 dQ \to \min \tag{7}$$

In SUePDF, the minimization of Eq. (7) is carried out in a loop procedure by varying the background reference and the fitting order $N$, which are initially input by users to optimize the normalized $I(Q)$. A normalized $I(Q)$ of a nanoporous carbon sample, scaled by the corresponding $\langle f_e^2(Q) \rangle = f_e^2(Q)$ is shown in Fig. (2).

## 2.3. Fourier Transform of $F(Q)$ to yield reduced PDF

The *reduced PDF* $G(r)$ is obtained by a Fourier transform of the *reduced structure function* $F(Q) = Q[S(Q) - 1]$:

$$G(r) = \frac{2}{\pi} \int_{Q_{\min}}^{Q_{\max}} F(Q) \sin(Qr) dQ \tag{8}$$

The red-dot curve shown in Fig. (3a) is the unfiltered $G(r)$ for nanoporous carbons after the Fourier transform [Eq. (8)]. Besides being convolved with the termination function $[\sin(Q_{\max}r) - \sin(Q_{\min}r)]/(\pi r)$ (Peterson et al., 2003), this unfiltered $G(r)$ exhibits some artifact peaks at low (< 1 Å) and high (> 20 Å) values of $r$, which are equivalent to the low- and high-frequency noises, respectively, of the experimental data.

## 2.4. PDF normalization and noise filtering

The unfiltered $G(r)$ obtained from Eq. (8) has to be adjusted by physical and mathematical constraints. These constraints refer back to the definition of PDF $g(r)$, which represents the probability density of finding a pair of two atoms separated by distance $r$ (Egami & Billinge, 2002). Therefore $g(r)$, the probability density, has to be non-negative:

$$g(r) = \gamma(r) + \frac{G(r)}{4\pi\rho_0 r} \geq 0 \tag{9}$$

where $\gamma(r)$ is the nanoparticle form factor (Kodama et al., 2006; Gilbert, 2008; Tran et al., 2016) [for bulk, $\gamma(r) \equiv 1$] and $\rho_0$ is the average number density of the sample. The normalization of the PDFs is based on Eq. (9). Prior knowledge about the shortest interatomic distance existing in the sample is used as a physical constraint for which the PDF $g(r)$ at distances smaller than a value $r_{min}$ is set to zero:

$$\begin{aligned} g(r) &= 0 \\ G(r) &= -4\pi\rho_0 r\gamma(r) \end{aligned} \quad \text{when} \quad r < r_{min} \quad (10)$$

Eq. (10) is used to filtering off low-frequency noises. The high-frequency noises can be filtered of by setting an upper cut-off distance $r_{max}$, where $G(r) = 0$ at $r > r_{max}$. This noise treatment of $G(r)$ is demonstrated in Fig. (3a). A Fourier back-transform of the treated $G(r)$ then yields the noise-filtered $F(Q)$ data, shown in Fig. (3b), for the here given example from nanoporous carbon sample.

## 2.5. Evaluation of uncertainty

The treated reduced PDF $G_T(r)$ in a comparison with $G_{BFT}(r)$ which is the Fourier transform of the [1.2-20 Å] band-pass filtered $F(Q)$, is shown in Fig. (4a). The observed difference between $G_T(r)$ and $G_{BFT}(r)$ is the consequence of the cutting of low and high frequencies which is propagated through the Fourier transform. Therefore, the uncertainty considered here is mainly associated with the low- and high-frequency noises, which correspond to the artifact short- and long-distance peaks found in the unfiltered $G(r)$, respectively.

Based on Fig. (3b), it is clear that the low-frequency noises can be attributed to some low-frequency distortions of the $F(Q)$ data. These distortions may be caused by the imperfectness of the instrumental setup, the errors encountered during the background subtraction and data scaling using the approximated electron scattering factors. One of typical errors is the beam convergence, which generally does not affect the positions of PDF peaks but does affect their intensities (McBride & Cockayne, 2003). Beside these errors, in the cases of coherent structures, the low-frequency distortions may be related to the coherent multiple scattering that might not be treated properly by background subtraction. The high-frequency noises, which are believed to be more random, may come from various sources. Some of probable sources are: electromagnetic environment, mechanical instability of the instrument, etc. It is worth to note that the properties of the recording media are of important and can influence the results in both images and diffraction significantly, in particular resolution and sensitive (Ruskin et al., 2013). In general, higher pixels in the detector better resolution can be achieved. The current generation of detectors, which is like CCD and CMOS-types and have either 1k x 1k or 2k x 2k pixels installed in a typical electron microscope configuration, can provide significantly high $Q_{max}$ for most applications. A detector with better dynamical range and sensitive can have the advantage to manage strong diffraction at low-Q range and to acquire the weaker signal at higher Q-range. As the results, the extended Q-range can improve the overall quality of ED-based PDF. It can be mentioned that the optimization of the acquisition condition for a particular detector can help to obtain quality electron diffraction patterns and therefore ED-based PDF, especially to minimize artefacts given by the detector, such as streaking, trace of beam blank, blooming, etc.

The uncertainty of the normalized PDFs can be evaluated based on the relative root-mean-square (*rms*) difference between the corresponding $g_T(r)$ and $g_{BFT}(r)$:

$$U_{g([r_{\min},r_{\max}])} = \frac{rms\{g_T(r) - g_{BFT}(r)\}}{rms\{g_T(r)\}} \tag{11}$$

For the case of nanoporous carbon sample shown in Fig. (4b), $U_{g([1.2Å, 20Å])} \approx 3.7\%$.

## 3. The graphic user interface (GUI) of SUePDF

### 3.1. Electron total scattering profile input

The input data for SUePDF v1.0 is electron total scattering intensity 1D profile. The input file must have a simple x-y two-column format. The x-column is the *s*-values (in Å$^{-1}$ when calibrated; noted that $Q = 2\pi s$) and the y-column is the corresponding intensity *I(Q)*. Multiple input files may be selected for integration of different diffraction data of the same sample and in the same experimental conditions. Multiple input files must be synchronized with the same format and the same data size. Figure 5 shows an overview of the GUI after inputting a data file using *INPUT BROWSER*.

## 3.2. Loading electron scattering factors

Databases of parameterization for both electron (Kirkland, 2010) and X-ray (Brown *et al*, 2006) scattering factors (or atomic form factors) are implemented in SUePDF. In cases of neutral atoms the electron database will be used, otherwise the X-ray database will be loaded in order to calculate the electron scattering factors for ions via Mott-Bethe formula (Eq. 4). Beside the electron energy (in keV), chemical information, including the elemental composition, molar ratios and valences, is required as inputs to load appropriate electron scattering factors for data scaling. Figure 6 shows the GUI panel for inputting chemical information, electron kinetic energy and calculation of electron scattering factors.

## 3.3. Background optimization

The background will be optimized based on user input of the following:
- Two points specifying the pre-peak background (corresponding to the lowest momentum transfer) and the tail.
- Number of middle background reference points: these points will be automatically positioned as initial conditions between the previous two points of the pre-peak and the tail. Their positions will vary along the curve of the raw scattering profile to optimize the background. It is noted that these reference points generally do not lie on the background, their distances to the background are refined while their positions varies. The typically recommended numbers of these reference points are 3 – 8. Larger numbers of reference points consumes more computation time.
- Maximum fitting order: typically recommended orders are 5 – 15. Larger fitting orders consume more computation time.

Figure 7 shows a background optimized in SUePDF GUI, together with the scaled *I(Q)* and *S(Q)*.

### 3.4. *S(Q)* correction and high-frequency (HF) noise filtering

SUePDF offers a routine for correction of *S(Q)* tail. This routine is optional and only recommended when a good enough solution for *S(Q)* scaling problem can not be found by background optimization. The correction procedure is based on calculation of the median curve of the SQ tail. There are two steps:

1. Specifying the to-be-corrected tail of *S(Q)*.

2. Tuning the order of the median fitting to achieve a corrected tail of *S(Q)*. The tuning may also be done automatically by pressing *Auto optimization* button.

HF noise filtering is recommended because ED data usually contain high-frequency noises, which are more visible at high Q values as spiky oscillation. The filter is based on forward-and-back Fourier transforms. A cut-off distance is required as a user input. The cut-off distance serves as the "highest frequency" allowed in the ED data and supposed to relate to the atomic structure.

### 3.5. Nanoparticle form factor

The nanoparticle form factor takes into account the size and shape of the sample and quantifies how much they affect the normalized PDF. For bulk sample this factor is unity. The SUePDF v.1.0 offers the calculations for four basic shapes: sphere, cuboctahedron, cube, and truncated cube (Tran *et al.*, 2016). When working with nanoparticles samples, user must input their size and shape to load appropriate form factor. Without loading this, the default form factor will be unity (for bulk). Figure 8 shows the GUI panel for calculation of nanoparticle form factor.

### 3.6. PDF renormalization

This renormalization is to achieve quantitative PDF *g(r)*. The renormalization procedure is based on:

- The non-negativity of $g(r)$ as probability density.

- The cutting off of short unphysical distances contaminated as low-frequency distortion in the data (low-frequency filtering). This requires a user input of a lower cut-off distance (based on general prior-knowledge of the shortest interatomic distance existing in the sample).

- The revision of number density: SUePDF is able to deduce a value of number density from the normalized ED data. If a better value of number density has been well known already, it should be used as the correction for the deduced value. In a good case of data processing of nanoscale samples, the deduced value can be very close to the generally accepted one.

Figure 9 shows the PDF renormalization by cutting off unphysical distances of low-frequency distortions and number density revision.

## 3.7. PDF quantification

As shown in Figure 10, coordination number can be measured by specifying an integration window for the relevant $g(r)$ peak. The background-subtracted and noise-filtered electron scattering profile (which should be the extracted kinematical scattering data) can be reconstructed back into a ring pattern, shown as the inset of Figure 10.

## 4. Examples

### 4.1. Gold nanoparticles

A SUePDF example of ~5 nm sized Au nanoparticles is shown in Figure 11. The particle is considered to be close to spherical shape (thus, the nanoparticle form factor of a 5 nm sphere was chosen) although some facets resembling cuboctahedral morphology can be seen in the high-resolution TEM image (inset of Fig. 11a). To obtain a proper ED data of Au nanoparticles, which are supported on an amorphous carbon film, an ED data of an equivalent blank carbon film was collected as the substrate reference. The PDF based on an ED data of $Q_{min}$ = 1.5 Å$^{-1}$ and $Q_{max}$ = 12.5 Å$^{-1}$ is shown in Figure 11b (red) in a comparison with the theoretical PDF of a 5 nm spherical model of Au perfect fcc nanoparticle (blue).

## 4.2. Amorphous silica

Figure 12 shows the examples of amorphous silica at thin and thick areas, in order to demonstrate the validity of the background modeling for samples with different thicknesses. These areas are shown in the TEM images (Fig. 12a&b) with the selected-area apertures (marked with dash-blue and solid-red circles for the thin and thick areas, respectively) defining the regions for ED acquisition. The corresponding (reduced) PDFs obtained using SUePDF are shown in Figure 12c. The PDF of the thin area (dash-blue) was obtained from an ED data of $Q_{min}$ ~ 0.4 Å$^{-1}$ and $Q_{max}$ ~ 12 Å$^{-1}$. The PDF of the thick area (solid-red) was obtained with the same $Q_{max}$ but a slightly higher $Q_{min}$ (0.55 Å$^{-1}$) in order to cut off the possible increase of inelastic scattering. It is noted that, because of the limited $Q_{max}$, the ED-based PDFs are generally more broaden compared with X-ray/neutron-based PDFs and the termination ripples may interfere significantly with some low and broaden peaks of amorphous materials (e.g., Si-Si peaks). The quantitative measurements of bond lengths and coordination number for both the thin and thick areas are listed in Table 1, together the reference data from neutron scattering (Keen & Dove, 1999) and molecular dynamic MD simulation of bulk amorphous silica (Hoang, 2007). The bond lengths found by SUePDF do not change significantly from the thin to the thick areas. On the other hand, the average coordination numbers (measured with a number density of 0.065 Å$^{-3}$) do vary from the thin to the thick region. Apart from the possible errors specified in 2.5, this reasonable variance could be attributed to the finite size and local effects when the ED data obtained from small amounts of sample in a TEM, which are not enough to be fully considered as bulk samples.

## 4.3. Metallic glass

Figure 13 shows an ED-based PDF of $Cu_{0.475}Zr_{0.475}Al_{0.05}$ metallic glass in comparison with the corresponding X-ray data (Kaban *et al*., 2015). The ED data has the $[Q_{min}, Q_{max}]_{ED}$ = [0.9, 12.2] Å$^{-1}$ while the X-ray data has the $[Q_{min}, Q_{max}]_X$ = [0.7, 21.1] Å$^{-1}$. It is noted that the difference in $Q$ range can cause different termination effect on the ED-based PDF and X-ray PDF. Beside this, the sample amount in a TEM-based ED experiment is much less than the amount in a X-ray experiment, suggesting that the information given by the ED-based PDF is more local than the information given by X-ray PDF.

## 5. Environment and distribution of SUePDF

SUePDF is written in MATLAB language and is compiled as a stand-alone GUI program, which only requires the free MATLAB Runtime R2015a (or newer) environment (ww.mathworks.com) installed on 64-bit PC-Windows platforms (Windows XP or newer is recommended). The installer (*SUePDF_Installer.exe* file) will automatically download (internet connection required) and install the MATLAB Runtime environment before installing SUePDF when executed.

SUePDF is distributed as free software for academic users, with an installer file and a manual document available for free download at https://osf.io/c2jq8/.

## 6. Summary

We have described the implementation of SUePDF, a GUI program dedicated to structural analysis based on electron diffraction data. SUePDF facilitates the TEM-based structural studies of amorphous materials and nanoparticles by converting the electron diffraction data in the reciprocal space into the quantitative PDFs in the direct space. SUePDF employs the scattering physics of electrons as well as the physical meaning of PDFs to achieve reliable data normalization. Noises are treated in SUePDF by band-pass Fourier filtering that also allows the evaluation of uncertainties caused by experimental conditions and data treatment procedures. Examples of using SUePDF to obtain quantitative PDFs of crystalline Au nanoparticles, amorphous silica and amorphous $Cu_{0.475}Zr_{0.475}Al_{0.05}$ metallic glass have been demonstrated.

**Acknowledgements**   The Knut and Alice Wallenberg (KAW) Foundation is acknowledged for providing the electron microscopy facilities and financial support under the project 3DEM-NATUR.

**Table 1** Bond lengths and coordination numbers of amorphous silica found by using SUePDF on the ED data obtained from thin and thick areas; these values are compared with the results from neutron (N) scattering (Keen & Dove, 1999) and molecular dynamics (MD) simulations (Hoang, 2007); noted that the first peak of the total PDF measures the average of (Si-O) and (O-Si) coordination number which is calculated as $[N^{Coord}_{(Si-O)}+2N^{Coord}_{(O-Si)}]/3$.

|  | **Bond lengths (Å)** | | | **Coordination Number** | | |
| --- | --- | --- | --- | --- | --- | --- |
|  | Si-O | O-O | Si-Si | (Si-O) &(O-Si) | O-O | Si-Si |
| SUePDF Thin area | 1.63 | 2.62 | 3.40 | 2.49 | 5.49 | 3.89 |
| SUePDF Thick area | 1.63 | 2.64 | 3.41 | 2.69 | 5.18 | 4.44 |
| Refs. | 1.62 (N) | 2.63 (N) | 3.10 (N) | 2.66 (MD) | 6.07 (MD) | 3.78 (MD) |

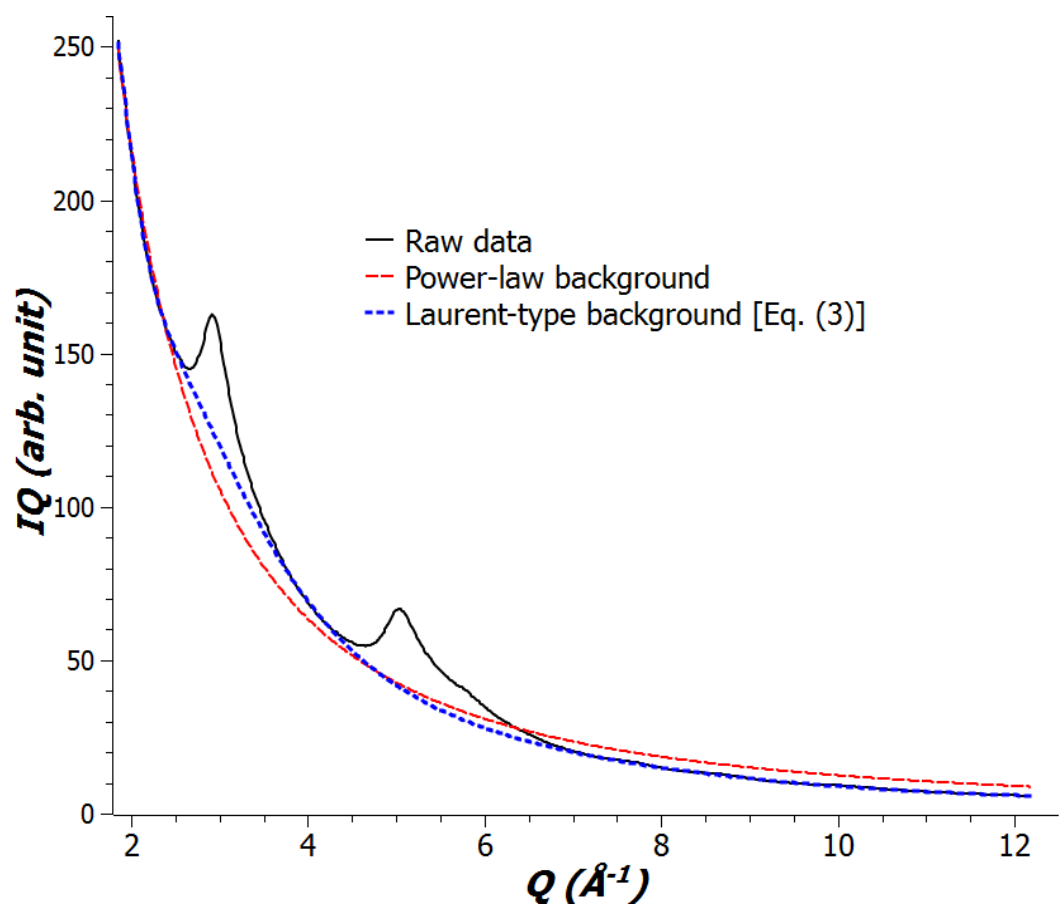

**Figure 1** Background modelling for electron powder diffraction data (solid-black) of nanoporous carbon: power-law model (dash-red) compared with Eq. (4) Laurent-type model (dot-blue) with $N = 7$.

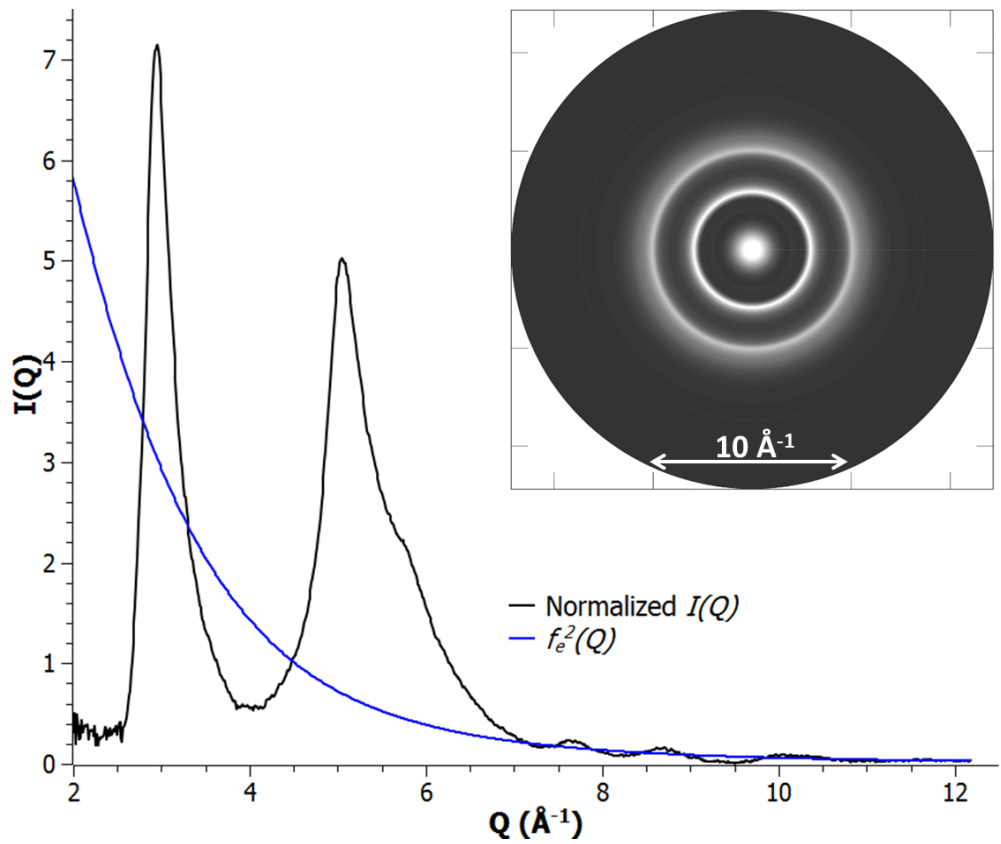

**Figure 2** Normalized $I(Q)$ of a nanoporous carbon sample (back solid) and the corresponding $\langle f_e^2(Q) \rangle = f_e^2(Q)$; the inset shows a ring pattern reconstructed from this normalized $I(Q)$.

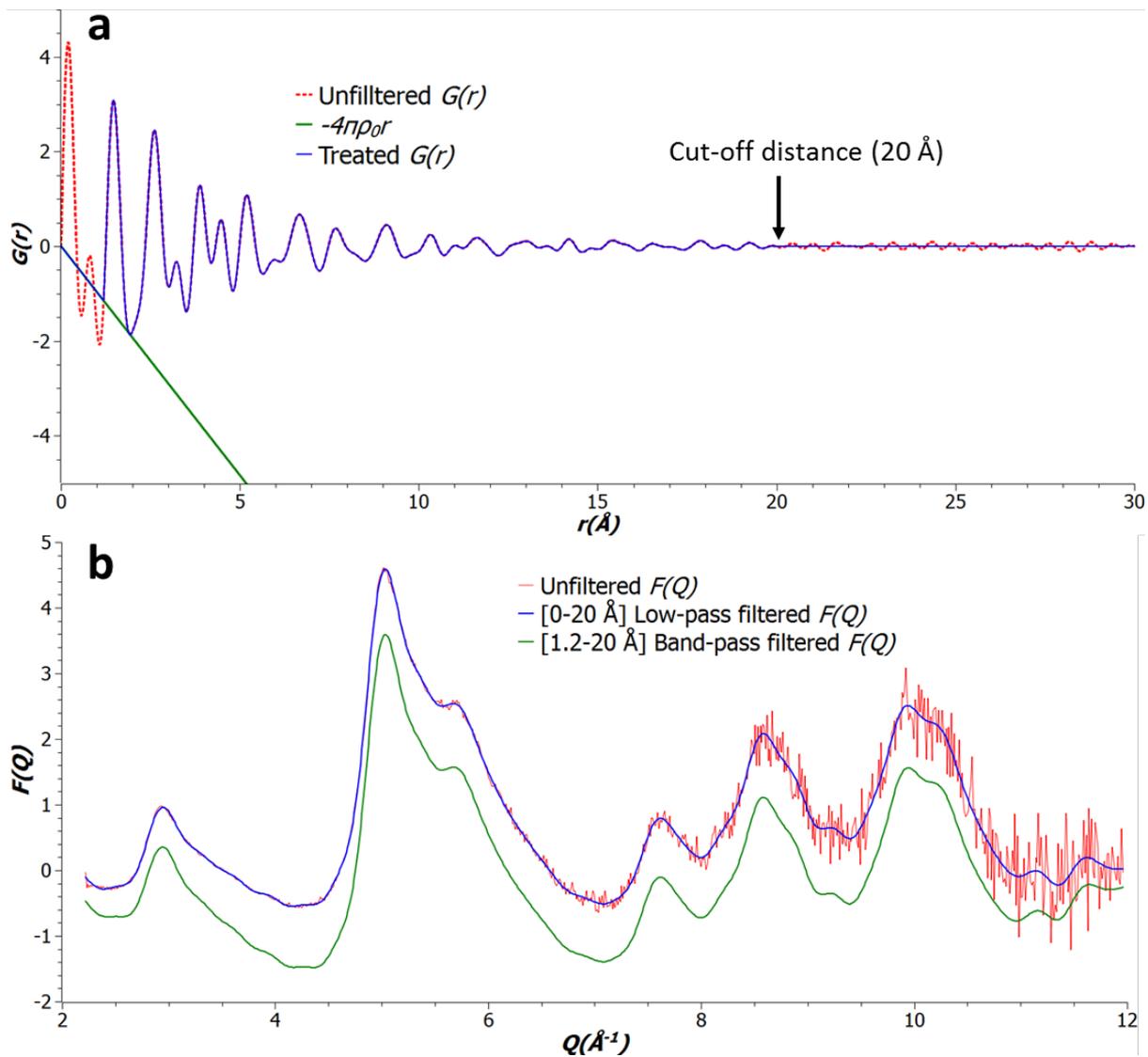

**Figure 3** a) Normalization and noise treating for $G(r)$ of nanoporous carbon, with $r_{min} \approx 1.2 \text{Å}$ and $r_{max} \approx 20 \text{Å}$; b) [0-20 Å] low-pass filtered $F(Q)$ (blue-solid) and [1.2-20 Å] band-pass filtered $F(Q)$ (green-solid) compared with the unfiltered $F(Q)$.

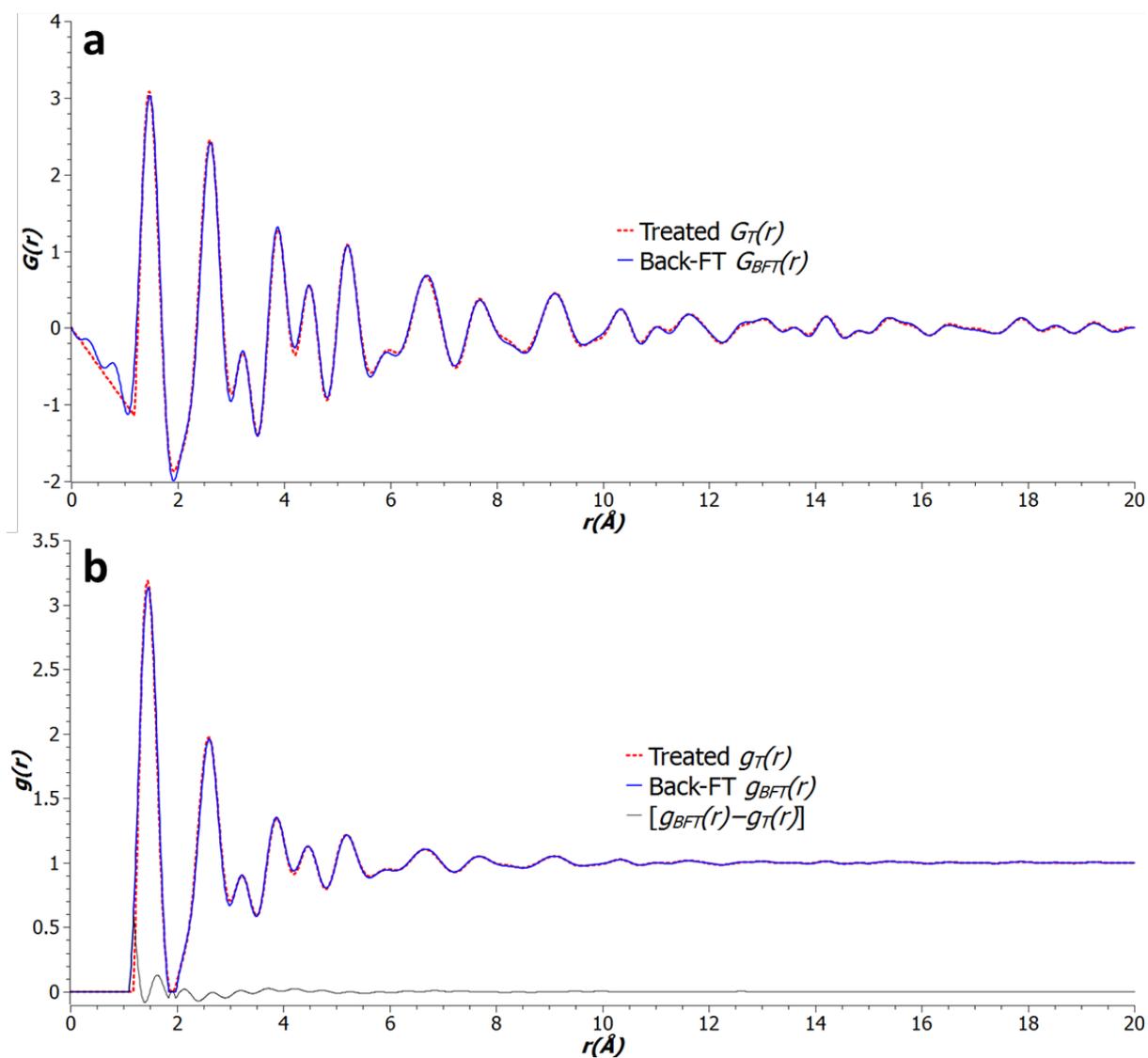

**Figure 4** Uncertainty in the noise filtering of PDFs: a) treated $G_T(r)$ [red-dot] compared with $G_{BFT}(r)$ [blue-solid] which has been transformed from [1.2-20 Å] band-pass filtered $F(Q)$; b) the corresponding normalized PDFs $g_T(r)$ [red-dot], $g_{BFT}(r)$ [blue-solid] and the difference between these (black-dot). Based on these, the evaluated uncertainty is ~3.7%.

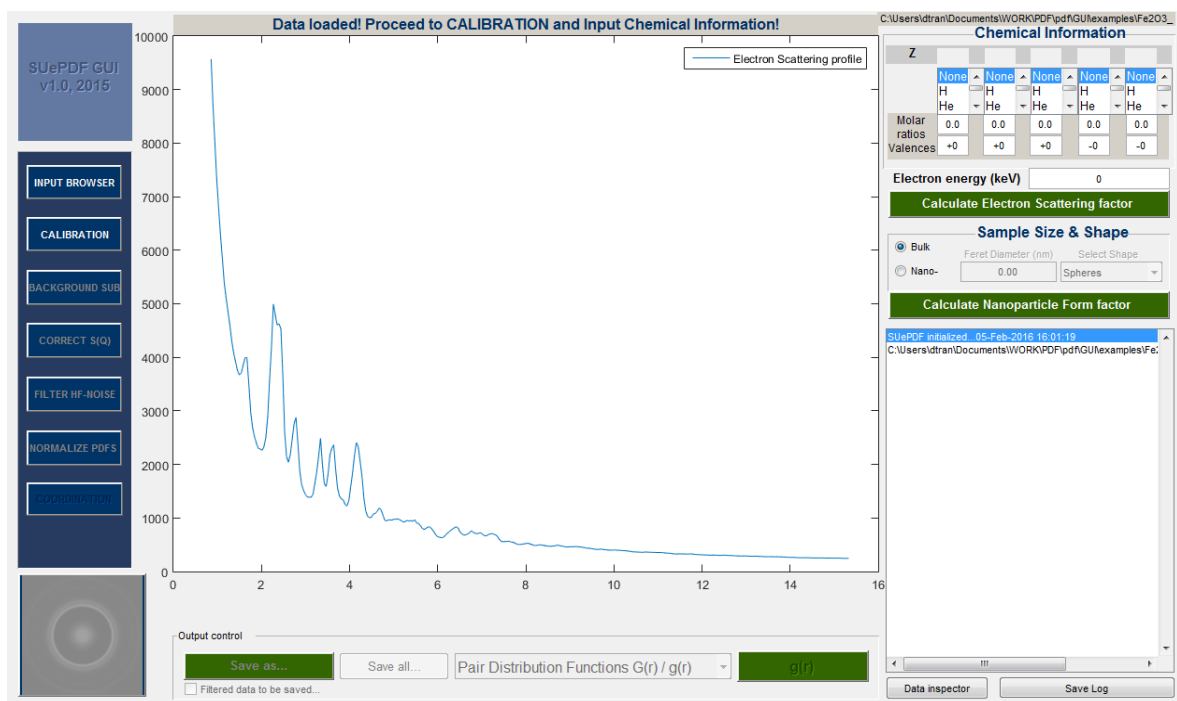

**Figure 5** An overview of the SUePDF GUI after inputting a data file of electron scattering profile.

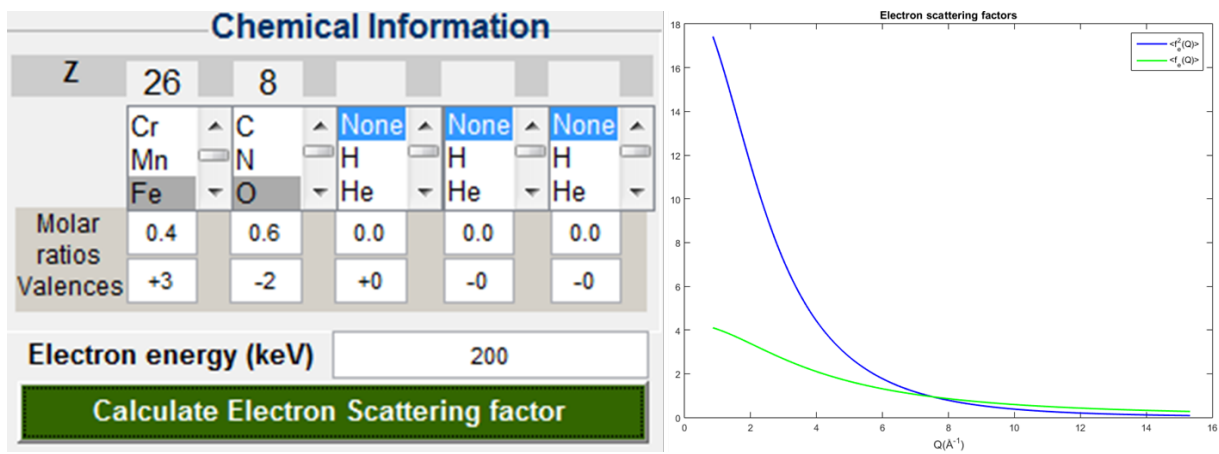

**Figure 6** The GUI for loading electron scattering factor by inputting chemical information and electron kinetic energy.

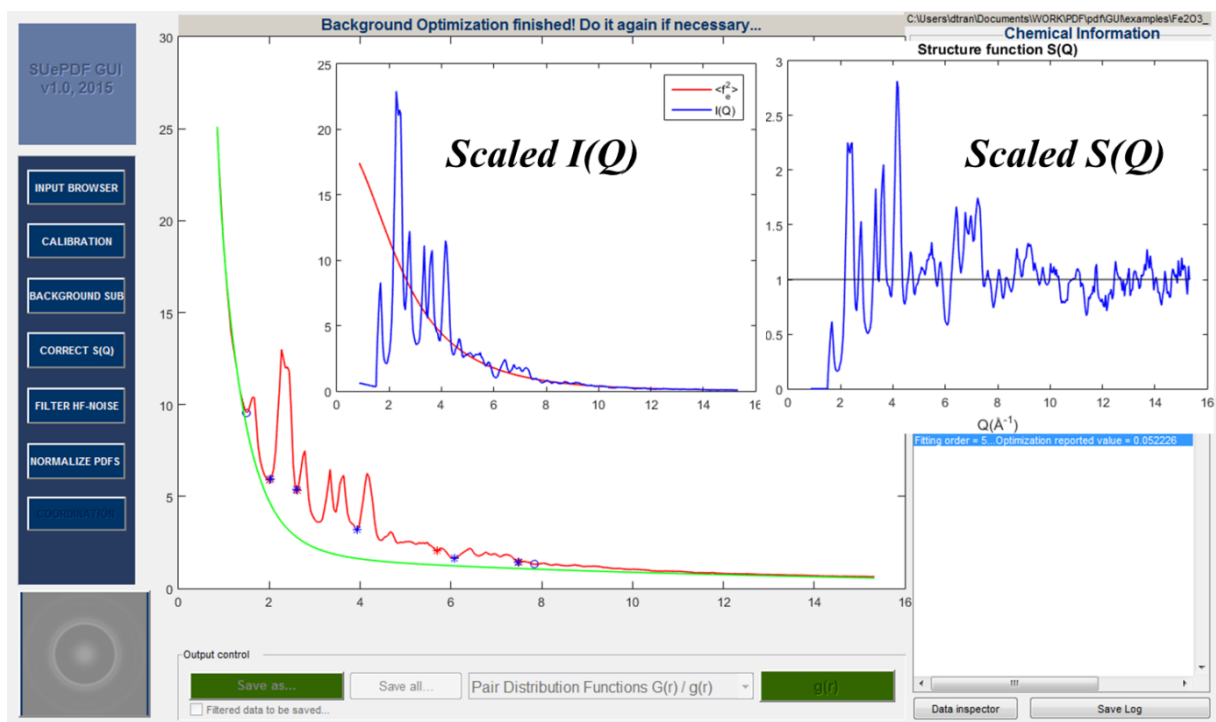

**Figure 7** An optimized background and the corresponding scaled *I(Q)* and *S(Q)*.

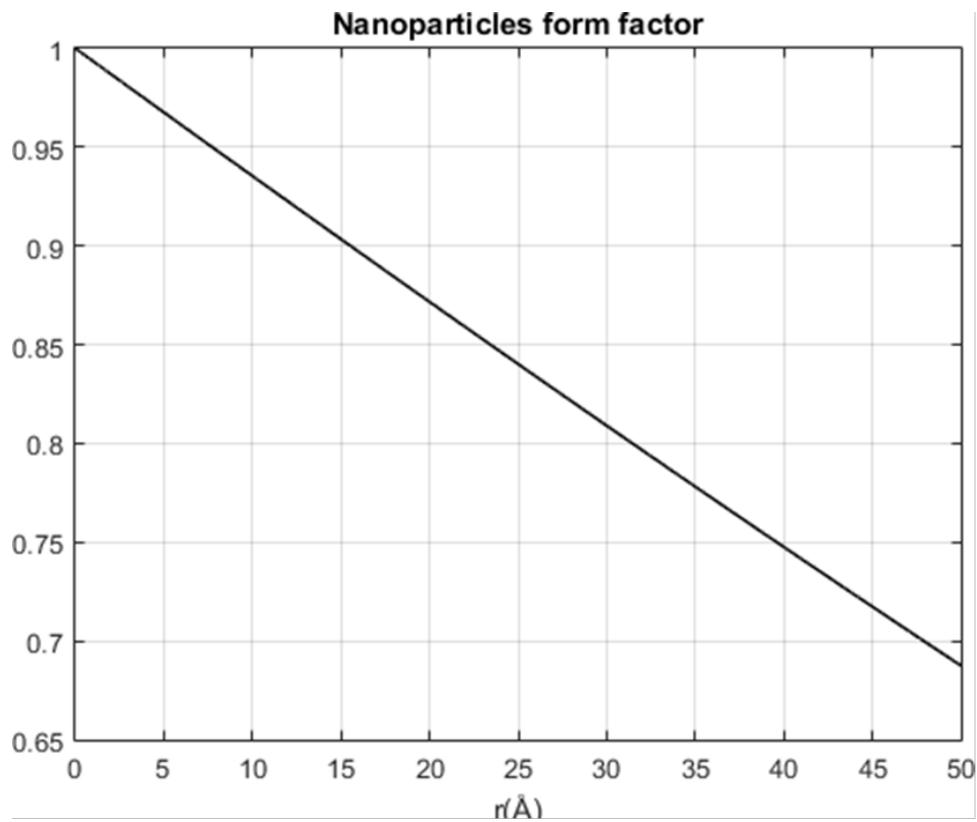

**Figure 8** GUI panel for calculation of nanoparticle form factors.

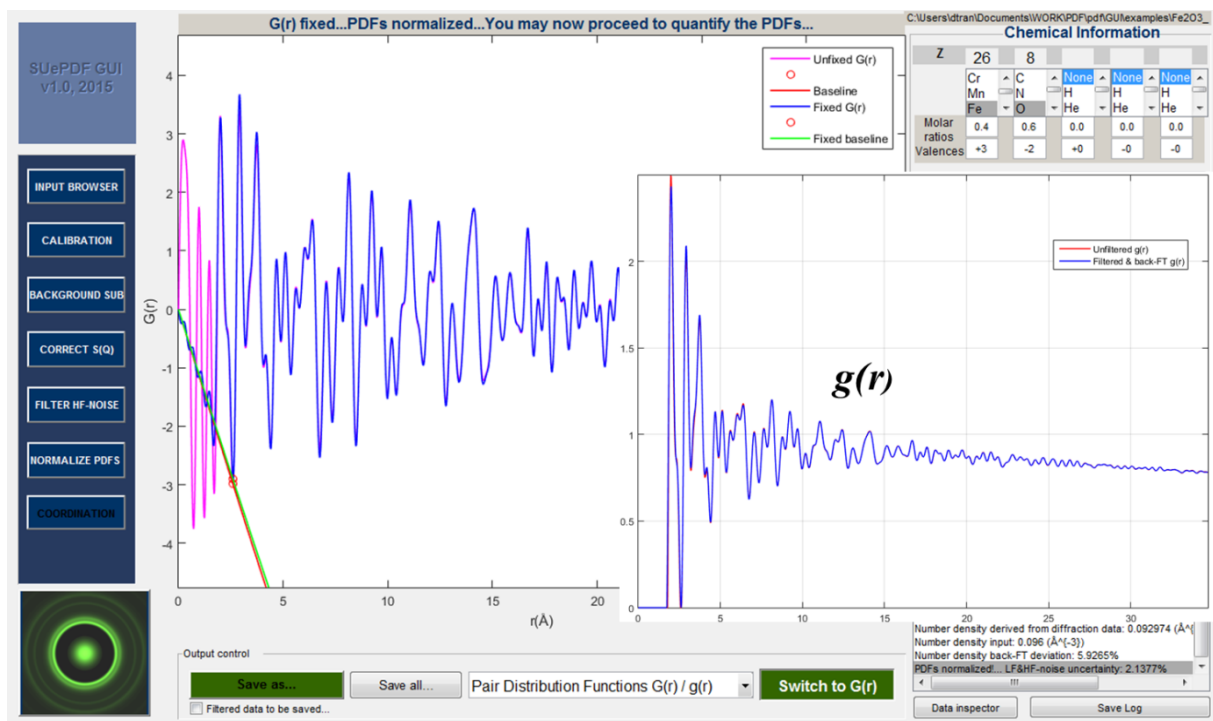

**Figure 9** PDF renormalization: the cutting off of low-frequency distortion and number density revision; $g(r)$ is shown in the inset.

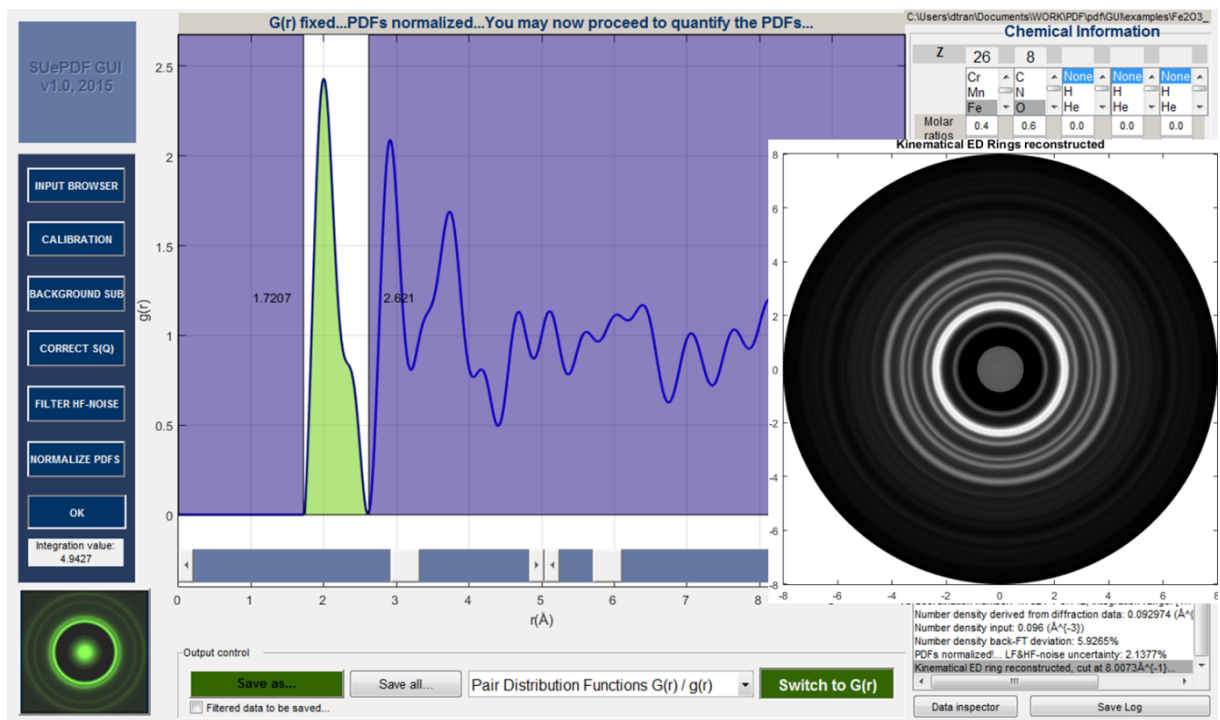

**Figure 10**    Measuring coordination number and reconstruction of kinematical ED pattern.

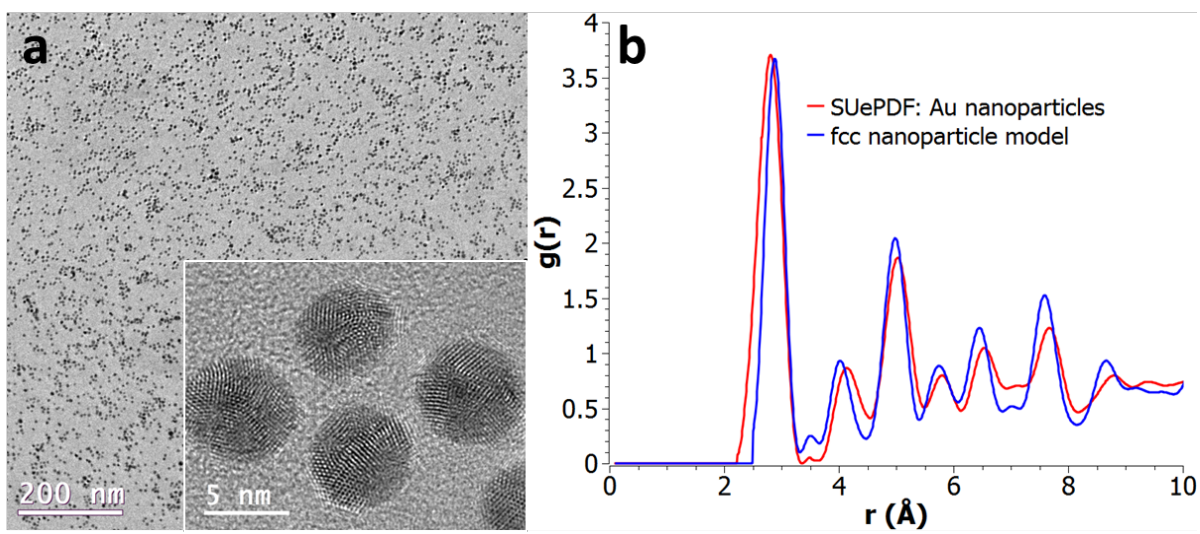

**Figure 11** (a) TEM images of ~5 nm –sized Au nanoparticles including a high-resolution image (inset); (b) ED-based PDF obtained using SUePDF (red) compared with the theoretical PDF of a 5 nm spherical Au nanoparticle model having perfect fcc structure (blue).

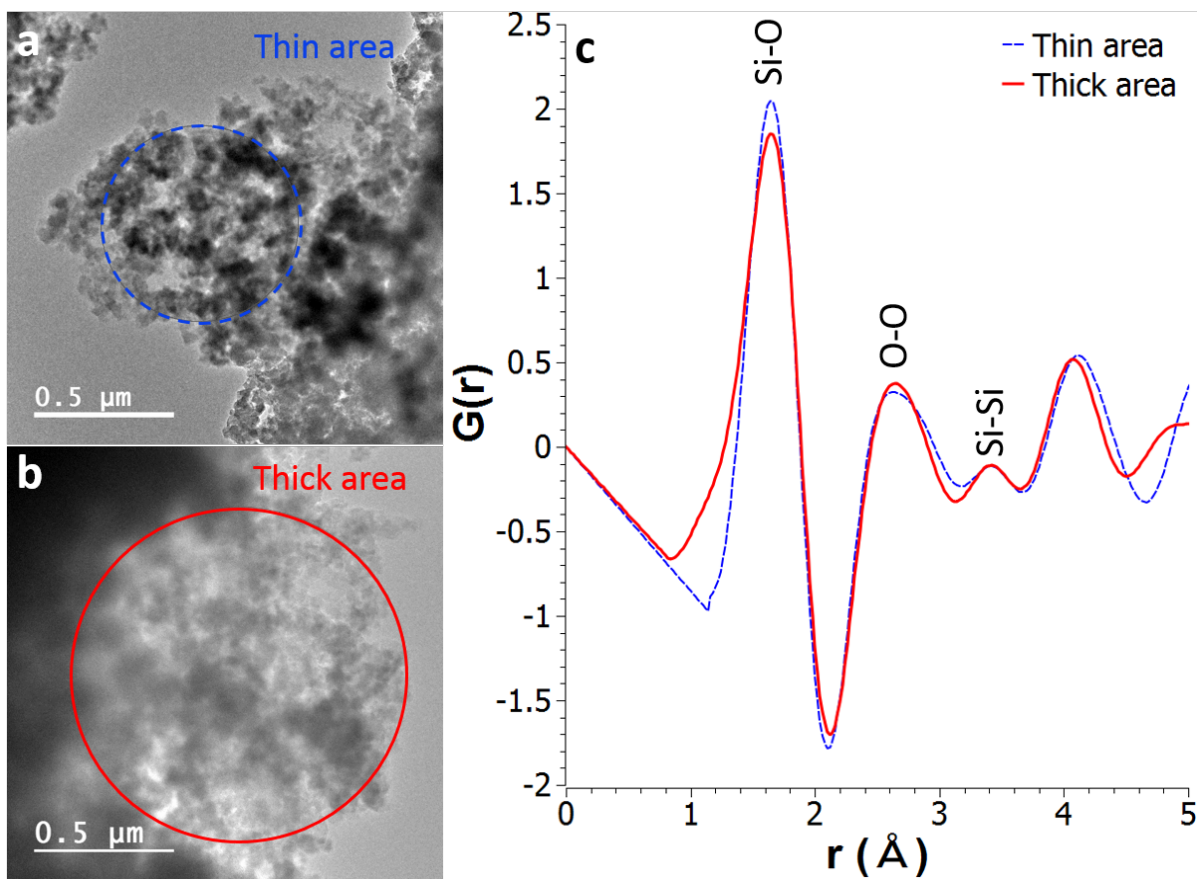

**Figure 12** Comparison between thin and thick areas of amorphous silica for obtaining ED-based PDF using SUePDF; (a) & (b) TEM images of thin and thick areas, respectively, for ED acquisition; (c) the corresponding ED-based PDFs for the thin (dash-blue) and thick (solid-red) areas.

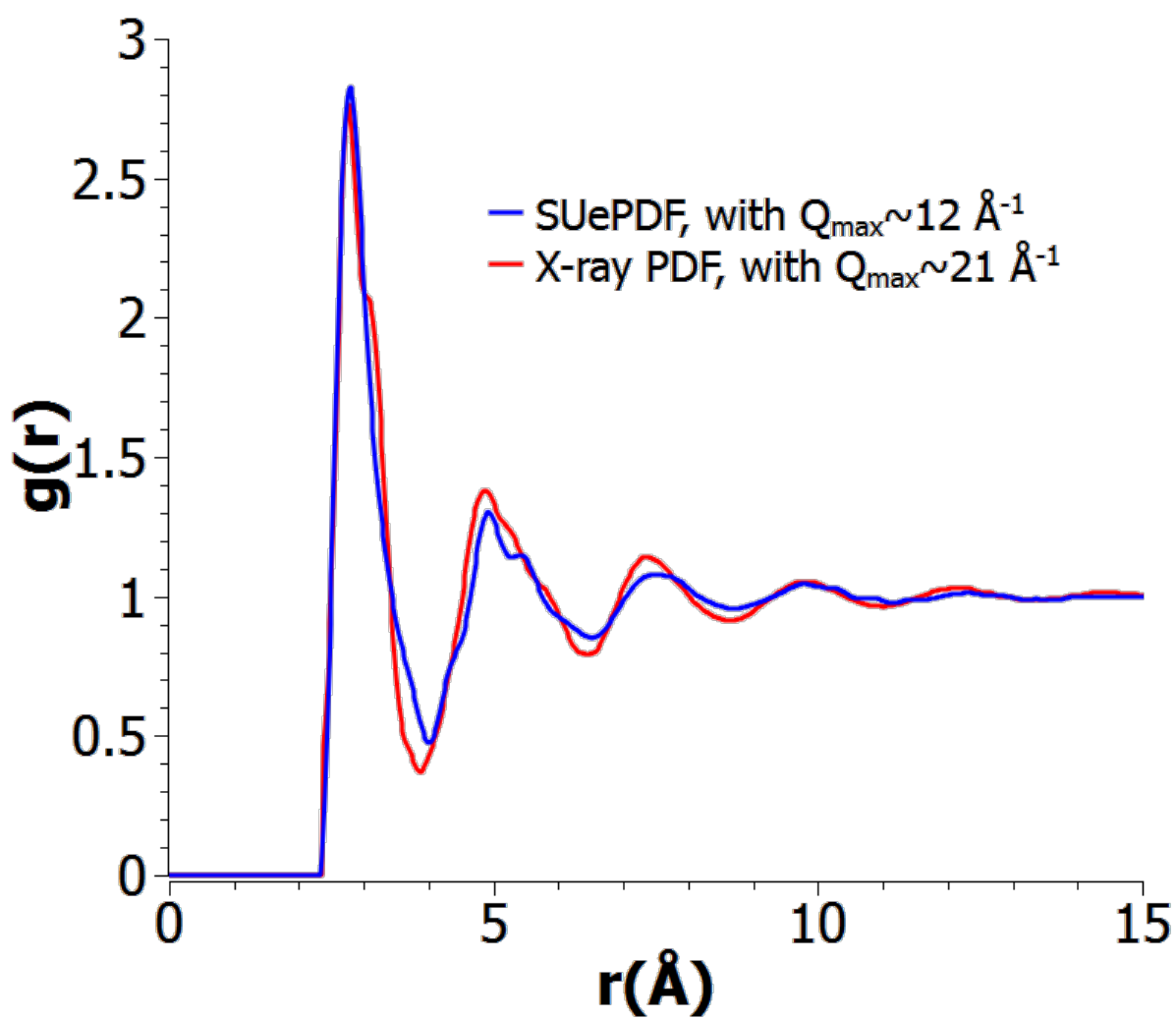

**Figure 13**  ED-based PDF (blue) of $Cu_{0.475}Zr_{0.475}Al_{0.05}$ metallic glass obtained using SUePDF compared with the corresponding X-ray data (Kaban *et al.*, 2015) shown in red.